\documentclass[twocolumn,showpacs,preprintnumbers,amsmath,aps]{revtex4}
\usepackage{graphicx}
\usepackage{dcolumn}
\usepackage{bm}
\usepackage{color}
\begin{document}
\def\eq#1{(\ref{#1})}
\def\fig#1{Fig.\hspace{1mm}\ref{#1}}
\def\tab#1{\hspace{1mm}\ref{#1}}
\title{Description of the thermodynamic properties of $\rm{BiH_{5}}$ and $\rm{BiH_{6}}$\\ 
superconductors beyond the mean-field approximation}  
\author{M. W. Jarosik $^{\left(1\right)}$}
\author{E. A. Drzazga $^{\left(1\right)}$}
\email{edrzazga@wip.pcz.pl}
\author{I. A. Domagalska $^{\left(1\right)}$}
\author{K. M. Szcz{\c{e}}{\'s}niak $^{\left(2\right)}$}
\author{U. St{\c{e}}pie{\'n} $^{\left(1\right)}$}
\affiliation{$^{\left(1\right)}$ Institute of Physics, Cz{\c{e}}stochowa University of Technology, Ave. Armii Krajowej 19, 42-200 Cz{\c{e}}stochowa, Poland}
\affiliation{$^{\left(2\right)}$ Ul. Pomorska 37/55, Zawiercie, 42-400, Poland}
\date{\today}
\begin{abstract}
The {\it ab initio} calculations suggest (Y. Ma {\it et al.}), that the high-pressure ($p=200$~GPa) superconducting state in 
$\rm{BiH_{5}}$ and $\rm{BiH_{6}}$ compounds characterizes with a high value of the critical temperature ($T_{C}\sim 100$~K). Due to the large value of the electron-phonon coupling constant ($\lambda\sim 1.2$), the thermodynamic parameters of the superconducting phase in $\rm{BiH_{5}}$ and $\rm{BiH_{6}}$ have been determined beyond the mean-field approximation - in the framework of the Eliashberg equations formalism. We have calculated the dependence of the order parameter ($\Delta$) and the wave function renormalization factor on the temperature. Then we have estimated the free energy difference between the superconducting state and the normal state, the thermodynamic critical field ($H_{C}$) and the specific heat of the superconducting state ($C^{S}$) and the normal state ($C^{N}$). The values of the dimensionless ratios $R_{\Delta}=2\Delta\left(0\right)/k_{B}T_{C}$, 
$R_{C}=\Delta C\left(T_{C}\right)/C^{N}\left(T_{C}\right)$ and $R_{H}=T_{C}C^{N}\left(T_{C}\right)/H^{2}_{C}\left(0\right)$ are equal 
$R_{\Delta_{{\rm BiH_{5}}}}=4.17$ and $R_{\Delta_{{\rm BiH_{6}}}}=4.20$, $R_{C_{{\rm BiH_{5}}}}=2.54$ and $R_{C_{{\rm BiH_{6}}}}=2.58$, 
$R_{H_{{\rm BiH_{5}}}}=0.146$ and $R_{H_{{\rm BiH_{6}}}}=0.146$ respectively.   
\end{abstract}
\pacs{74.20.Fg, 74.25.Bt, 74.62.Fj}
\maketitle
{\bf Keywords:} Hydrogen-rich materials, High-pressure effect, Electron-phonon superconductivity, Eliashberg theory, Thermodynamic properties
\vspace*{0.5cm}

In 1935 Wigner and Huntington noted that hydrogen affected by the high pressure ($\sim 25$~GPa) should transit into a metallic state \cite{Wigner1935A}. So far, it has been possible to experimentally demonstrate the metallization of liquid hydrogen for the pressure at $140$~GPa, but the temperature was up to $3000$~K \cite{Weir1996A}. In 2017 the information was spread that hydrogen metallization was observed at room temperature under the pressure at $495$~GPa \cite{Dias2017A}. Unfortunately, this result has not been confirmed again.

In 1968 Ashcroft noted that the metallic hydrogen should be a high-temperature superconductor \cite{Ashcroft1968A}. He gave the following arguments: first of all, hydrogen has the lowest atomic mass of the nucleus, which implies the Debye frequency ($\omega_{D}$) of the crystal lattice of solidified hydrogen to be high; secondly, there are no electrons on the internal shells in the hydrogen atom, which favors a strong electron-phonon coupling. In the following years, thanks to the rapid development of computer technology, more detailed calculations were carried out. The hydrogen metallization pressure was estimated to be around $400$~GPa \cite{Ashcroft1968A, Stadele2000A}. It has been suggested that in the molecular hydrogen phase, within the range of pressures from $400$-$500$~~GPa  there should be a sharp rise in the critical temperature from approximately $80$~K to about $350$~K \cite{Cudazzo2008A, Szczesniak2013C, Yan2011A, Szczesniak2012N}. Above the pressure at $500$~GPa the numerical calculations suggest that the metallic molecular hydrogen should dissociate into the atomic phase \cite{Cudazzo2008A}, whereas the critical temperature value should be within the limit from $300$~K to $470$~K \cite{Maksimov2001A}. In the literature on the subject there are also works analyzing the superconducting state of hydrogen for extremely high pressures (up to $3.5$~TPa) \cite{Maksimov2001A, Szczesniak2009A, Szczesniak2014F, McMahon2011A, McMahon2011B}. The highest value of $T_{C}$ was foreseen at $630$~K for $p=2$~TPa. 

Due to the very high value of the hydrogen's metallization pressure, in 2004 Ashcroft suggested to try to look for a high temperature superconducting state not in pure hydrogen but in hydrogenated compounds \cite{Ashcroft2004A}. 
Induction of high temperature superconducting state in hydrogenated compounds should be supported by chemical precompression caused by heavier elements. The real breakthrough occurred only in December 2014, when Drozdov, Eremets, and Troyan have experimentally demonstrated the existence of a high-temperature superconducting state in $\rm{H_{3}S}$ ($T_{C}=203$ K for $p=155$ GPa) \cite{Drozdov2014A, Drozdov2015A}. 

In the presented work, we have determined the thermodynamic properties of the superconducting state in hydrogenated compounds 
$\rm{BiH_{5}}$ and $\rm{BiH_{6}}$. Due to the high predicted value of the electron-phonon coupling constant ($\lambda_{\rm BiH_{5}}=1.236$ and $\lambda_{\rm BiH_{6}}=1.259$ \cite{Ma2015A}) we performed the calculations for both cases as a part of Eliashberg formalism \cite{Eliashberg1960A}.  A detailed description of Eliashberg equations and the computational methods used has been presented in the works \cite{Domin2014A, Domin2015A, Domin2016A, Domin2017A, Ania2014D, Ania2015A, Ania2015B}. The spectral function ($\alpha^{2}F\left(\Omega\right)$) used in numerical calculations has been determined by Y. Ma {\it et al.} with an use of the density functional theory \cite{Ma2015A}. Additionally, $1100$  Matsubara frequencies were taken into account, and the value of Coulomb pseudo-potential ($\mu^{\star}$) equals $0.1$. was adopted. The stability of the Eliashberg equations' solutions was obtained for a temperature higher or equal to $T_{0}=20$~K.

\begin{figure}
\includegraphics[width=\columnwidth]{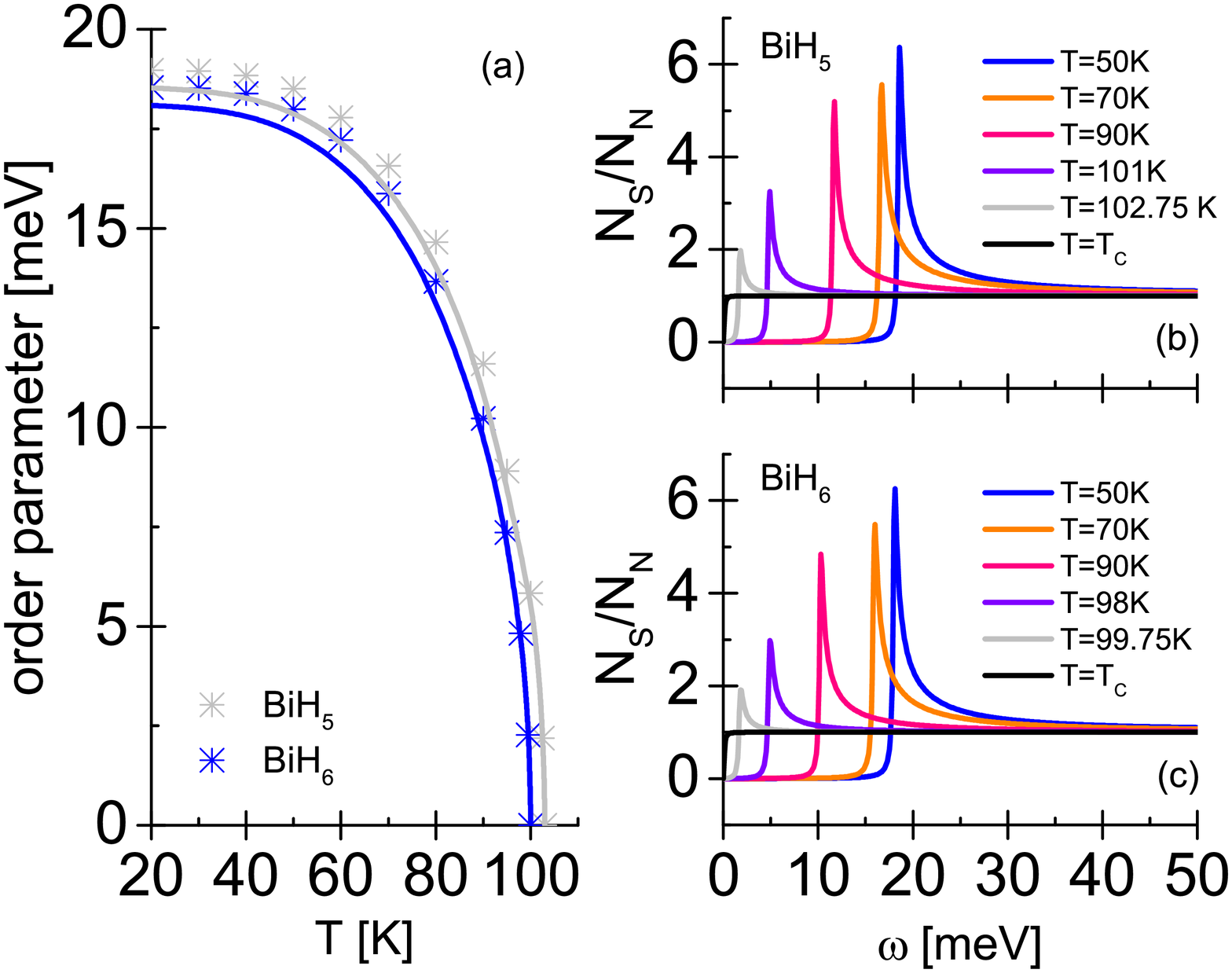}
\caption{(a) The order parameter as a function of the temperature obtained in the framework of the Eliashberg equations on the imaginary axis (lines). Symbols represent the values of the order parameter obtained on the basis of the equation \eq{r1}. 
         (b)-(c) Renormalized density of states for the selected values of temperature.
} 
\label{f1}
\end{figure}

In \fig{f1} (a) we have plotted a temperature dependence of the order parameter. It was obtained as part of the imaginary axis formalism and using the equation:
\begin{equation}
\label{r1}
\Delta(T)={\rm Re}[\Delta(\omega=\Delta(T))],
\end{equation}
where function of the order parameter on the real axis ($\Delta\left(\omega\right)$) was obtained using the method of the analytical extension, which has been described in detail in the paper \cite{Beach2000A}. We have obtained the following critical temperature values: $\rm{T_{C_{\rm{BiH_{5}}}}}=103$~K and $\rm{T_{C_{\rm{BiH_{6}}}}}=100$~K. It means that $\rm{BiH_{5}}$ and $\rm{BiH_{6}}$ compounds should belong to the family of high-temperature superconductors.

Let us consider the order parameter $\Delta(T_{0})=\Delta\left(0\right)$ estimated on the basis of the equation \eq{r1}. 
Hence the dimensionless ratio $R_{\Delta}=2\Delta\left(0\right)\slash k_{B}T_{C}$ equals $4.17$ and $4.2$, for $\rm{BiH_{5}}$ and 
$\rm{BiH_{6}}$. respectively. Compared with the predictions of the BCS theory, these are very high values, because $R_{\Delta_{\rm BCS}}=3.53$ 
\cite{Bardeen1957A, Bardeen1957B}. Please note that the above result comes from the existence of the significant retardation and strong-coupling effects. They can be characterized by the parameter $r=k_{B}T_{C}\slash\omega_{{\rm \ln}}$, where the quantity $\omega_{{\rm ln}}=\exp\left[\frac{2}{\lambda}\int^{\Omega_{\rm{max}}}_{0}d\Omega\frac{\alpha^{2}F\left(\Omega\right)}
{\Omega}\ln\left(\Omega\right)\right]$ denotes the logarithmic phonon frequency. For $\rm{BiH_{5}}$ and $\rm{BiH_{6}}$ compounds it was obtained $r=0.11$. In the BCS limit we have $r=0$ \cite{Carbotte2003A}. 

Basing on the function of the order parameter on the real axis, one can determine the renormalized electronic density of states in the superconducting phase:
\begin{eqnarray}
\label{r2}
\frac{N_{S}(\omega)}{N_{N}(\omega)}={\rm Re}\left[\frac{(\omega-i\Gamma)}{\sqrt{(\omega-i\Gamma)^2-(\Delta(\omega))^2}}\right],
\end{eqnarray}
where we have assumed that the pair breaking parameter equals $\Gamma=0.15$ meV. We have plotted the obtained results in the drawings \fig{f1} (b) and (c). 
Note that the characteristic maxima of the determined curves are formed in points $\omega=\pm\Delta$. On their basis, one can experimentally analyze the evolution of the energy gap depending on the temperature.

\begin{figure*}
\includegraphics[scale=0.4]{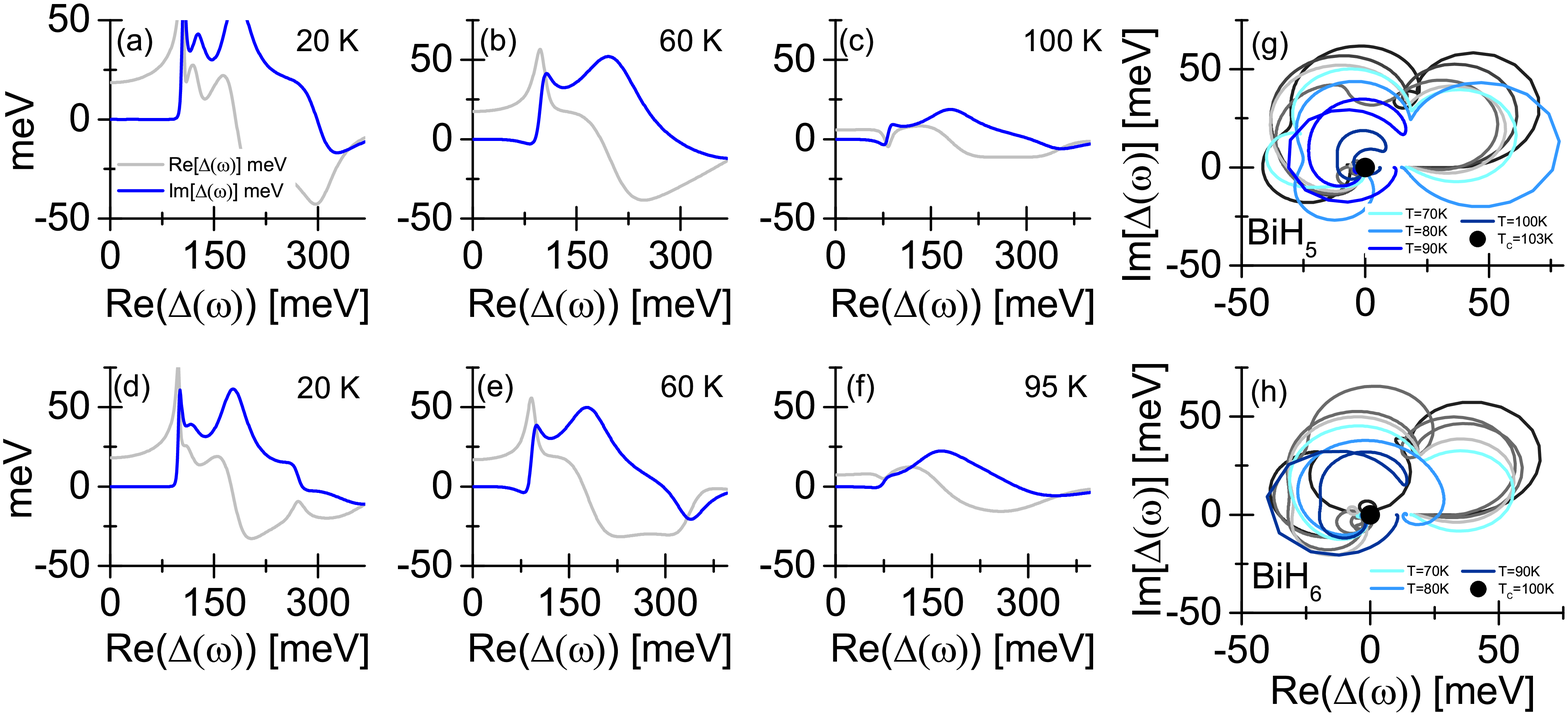}
\caption{(a)-(f) The real and imaginary parts of the order parameter on the real axis for the selected temperature values. (g) and (h) The values of the order parameter on the complex plane. The curves correspond to the frequency range from $0$ to $\omega_{D}$.} 
\label{f2}
\end{figure*}

The exact form of the order parameter on the real axis is presented in the figures \fig{f2} (a)-(f). 
When analyzing the case $T=T_{0}$, it can be seen that in the range of the lower frequencies ($\omega<\omega_{d}=85$~meV) non-zero is only the real part of the order parameter. This means no damping processes that are modeled by a function  ${\rm Im}\left[\Delta\left(\omega\right)\right]$. 
It is also worth noting that as the temperature rises, $\omega_{d}$ decreases significantly. Additionally, for illustrative purposes in drawings (g) and (h) we have plotted the values of the order parameter on the complex plane. In both analyzed cases, the values of $\Delta(\omega)$ on the complex plane form characteristic spirals, whose radii decrease with increasing temperature.

\begin{figure}
\includegraphics[scale=0.3]{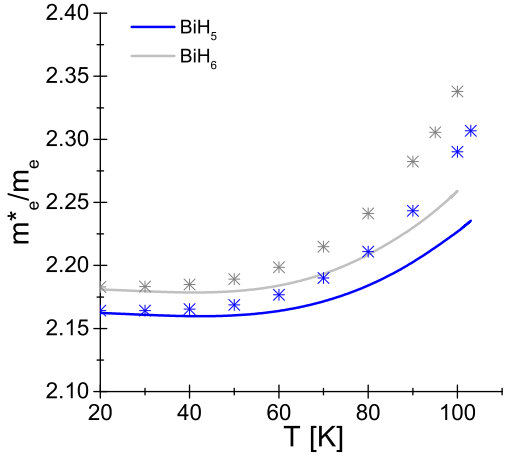}
\caption{The ratio of the electron effective mass to the electron band mass determined for $\rm{BiH_{5}}$ and $\rm{BiH_{6}}$ compounds as a part of the Eliashberg formalism of the imaginary axis (lines) and with the aid of formula \eq{r3} (stars).}  
\label{f3}
\end{figure}

The maximum value of the wave function renormalization factor on the imaginary axis relatively precisely determines the ratio of the effective mass of the electron ($m^{\star}_{e}$) to the electron band mass ($m_{e}$). However, the exact value of $m^{\star}_{e}\slash m_{e}$ should be determined on the basis of the following formula: 
\begin{equation}
\label{r3}
\frac{m^{\star}_{e}}{m_{e}}=Z\left(\omega=0\right).
\end{equation}

Graphs of the ratio of the electron effective mass to electron band mass in the temperature range from $0$ to $T_{C}$, were collected in the drawing \fig{r3}. It can be seen that in both cases the effective mass takes values slightly higher than two. This result proves a strong renormalization of the electron band mass by the electron-phonon interaction. Additionally, the ratio $m^{\star}_{e}\slash m_{e}$ is slightly dependent on the temperature, and its value at the critical temperature can be calculated with good accuracy using the formula: 
$\left[m^{\star}_{e}\slash m_{e}\right]_{T=T_{C}}=1+\lambda$ \cite{Carbotte2003A}. Let us note that for $T=T_{C}$ the numerical calculations give 
$\left[m^{\star}_{e}\slash m_{e}\right]_{{\rm BiH_{5}}}=2.3$ and $\left[m^{\star}_{e}\slash m_{e}\right]_{{\rm BiH_{6}}}=2.33$, which is in a good agreement with the analytical result. 

\begin{figure}
\includegraphics[width=\columnwidth]{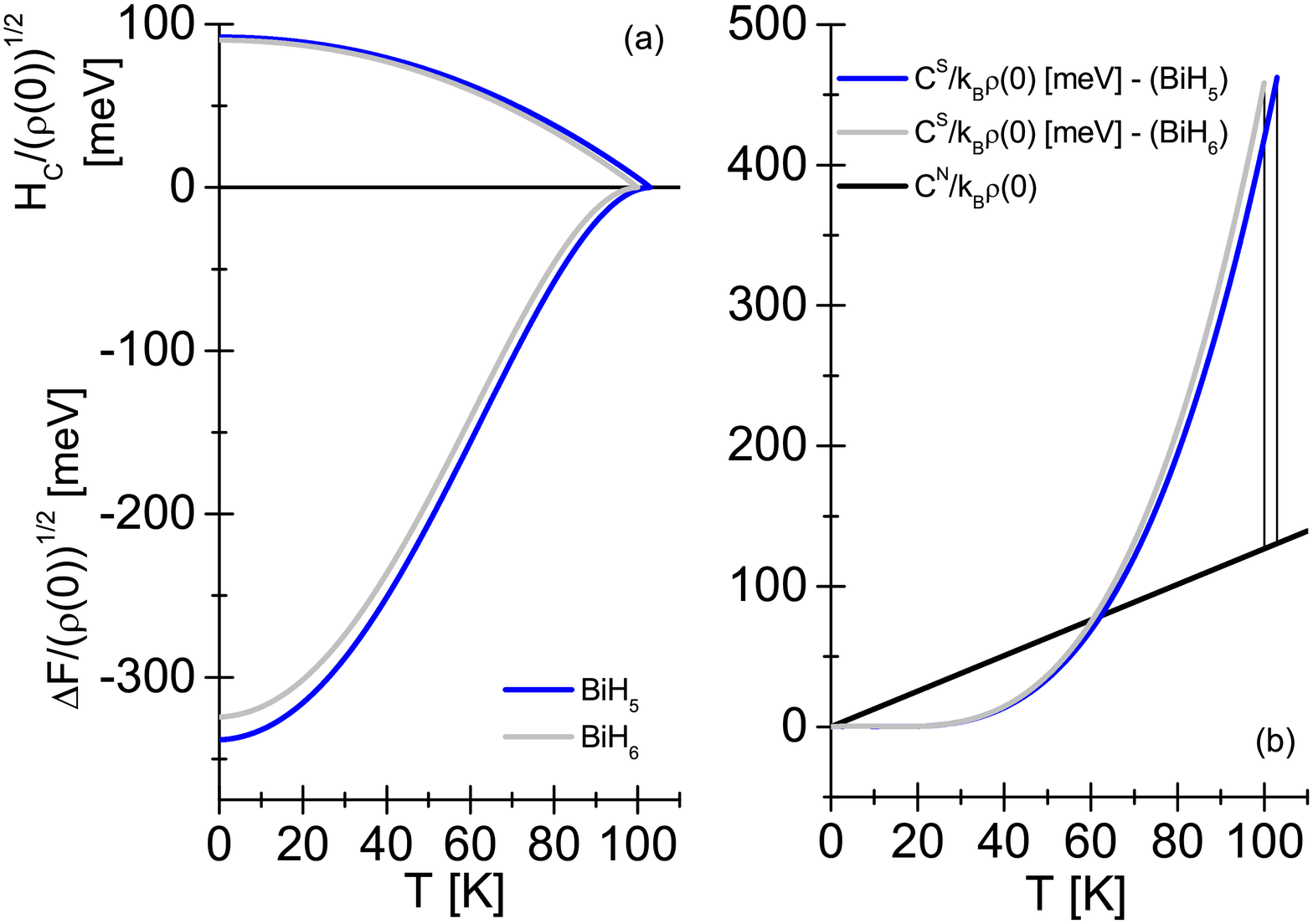}
\caption{(a) The free energy as a function of the temperature (lower panel) and the thermodynamic critical field (upper panel). (b) The specific heat of the superconducting state and the normal state as a function of the temperature.} 
\label{f4}
\end{figure}

 Basing on the solutions of the Eliashberg equations on the imaginary axis ($\Delta_{n}$ and $Z_{n}$) we have calculated the free energy difference between the superconducting state and the normal state: 
\begin{eqnarray}
\label{r4}
\frac{\Delta F}{\rho\left(0\right)}&=&-2\pi k_{B}T\sum_{n=1}^{M}
\left(\sqrt{\omega^{2}_{n}+\Delta^{2}_{n}}- \left|\omega_{n}\right|\right)\\ \nonumber
&\times&(Z^{S}_{n}-Z^{N}_{n}\frac{\left|\omega_{n}\right|}
{\sqrt{\omega^{2}_{n}+\Delta^{2}_{n}}}),
\end{eqnarray}  
where $\rho\left(0\right)$ denotes the value of the electron density of states at the Fermi level, $\omega_{n}$ is the fermion Matsubara frequency, and symbols $S$ and $N$ refer respectively to the superconducting and normal (metallic) state. On this basis, the thermodynamic critical field can be estimated:
\begin{equation}
\label{r5}
\frac{H_{C}}{\sqrt{\rho\left(0\right)}}=\sqrt{-8\pi\left[\Delta F/\rho\left(0\right)\right]}
\end{equation}
and the difference in specific heat between the superconducting state and the normal state ($\Delta C=C^{S}-C^{N}$): 
\begin{equation}
\label{r6}
\frac{\Delta C\left(T\right)}{k_{B}\rho\left(0\right)}=-\frac{1}{\beta}\frac{d^{2}\left[\Delta F/\rho\left(0\right)\right]}{d\left(k_{B}T\right)^{2}},
\end{equation}
wherein the specific heat of the normal state can be determined on the basis of the formula: 
$C^{N}(T)\slash{k_{B}\rho(0)}=\gamma k_{B}T$, where {$\gamma=\frac{2}{3}\pi^{2}(1+\lambda)$} denotes the Sommerfeld constant.

Obtained results were collected in the drawing \fig{f4}. It can be seen that the calculated thermodynamic functions assume similar values for $\rm{BiH_{5}}$ and $\rm{BiH_{6}}$, compounds, which results from similar values of the electron-phonon coupling constants. Additionally, it should be emphasized that the values of the thermodynamic critical field and specific heat cannot be correctly calculated within the framework of the mean-field BCS theory. To find out about this, let us estimate the dimensionless parameters 
$R_{C}=\Delta C\left(T_{C}\right)\slash C^{N}\left(T_{C}\right)$ and $R_{H}=T_{C}C^{N}\left(T_{C}\right)\slash H^{2}_{C}\left(0\right)$. For $\rm{BiH_{5}}$ compound we have received $R_{C}=2.54$ i $R_{H}=0.146$, while for $\rm{BiH_{6}}$ compound we have $R_{C}=2.58$ and $R_{H}=0.146$. Recall that the BCS theory gives accordingly $1.43$ and $0.168$ \cite{Bardeen1957A, Bardeen1957B}. 

\vspace*{0.5cm}

In summary, in the presented work we have determined the values of the thermodynamic parameters of the superconducting state in $\rm{BiH_{5}}$ and $\rm{BiH_{6}}$compounds. The systems were located under the pressure at $200$~GPa. We have shown that for the Coulomb pseudopotential's value of $0.1$ the superconducting state with a very high critical temperature of the order of $100$~K is induced in both studied compounds.
Additionally, within the framework of Eliashberg formalism, we have determined the values of: the energy gap, the electron effective mass, the thermodynamic critical field and the specific heat of the superconducting state. Performed calculations prove that the thermodynamic parameters of the superconducting state in $\rm{BiH_{5}}$ and $\rm{BiH_{6}}$ compounds cannot be calculated correctly in the framework of the mean-field BCS theory. 
\bibliography{Bibliography}

\begin{thebibliography}{30}
\expandafter\ifx\csname natexlab\endcsname\relax\def\natexlab#1{#1}\fi
\expandafter\ifx\csname bibnamefont\endcsname\relax
  \def\bibnamefont#1{#1}\fi
\expandafter\ifx\csname bibfnamefont\endcsname\relax
  \def\bibfnamefont#1{#1}\fi
\expandafter\ifx\csname citenamefont\endcsname\relax
  \def\citenamefont#1{#1}\fi
\expandafter\ifx\csname url\endcsname\relax
  \def\url#1{\texttt{#1}}\fi
\expandafter\ifx\csname urlprefix\endcsname\relax\def\urlprefix{URL }\fi
\providecommand{\bibinfo}[2]{#2}
\providecommand{\eprint}[2][]{\url{#2}}

\bibitem[{\citenamefont{Wigner and Huntington}(1935)}]{Wigner1935A}
\bibinfo{author}{\bibfnamefont{E.}~\bibnamefont{Wigner}} \bibnamefont{and}
  \bibinfo{author}{\bibfnamefont{H.~B.} \bibnamefont{Huntington}},
  \bibinfo{journal}{The Journal of Chemical Physics}
  \textbf{\bibinfo{volume}{3}}, \bibinfo{pages}{764} (\bibinfo{year}{1935}).

\bibitem[{\citenamefont{Weir et~al.}(1996)\citenamefont{Weir, Mitchell, and
  Nellis}}]{Weir1996A}
\bibinfo{author}{\bibfnamefont{S.~T.} \bibnamefont{Weir}},
  \bibinfo{author}{\bibfnamefont{A.~C.} \bibnamefont{Mitchell}},
  \bibnamefont{and} \bibinfo{author}{\bibfnamefont{W.~J.}
  \bibnamefont{Nellis}}, \bibinfo{journal}{Physical Review Letters}
  \textbf{\bibinfo{volume}{76}}, \bibinfo{pages}{1860} (\bibinfo{year}{1996}).

\bibitem[{\citenamefont{Dias and Silvera}(2017)}]{Dias2017A}
\bibinfo{author}{\bibfnamefont{R.~P.} \bibnamefont{Dias}} \bibnamefont{and}
  \bibinfo{author}{\bibfnamefont{I.~F.} \bibnamefont{Silvera}},
  \bibinfo{journal}{Science} \textbf{\bibinfo{volume}{355}},
  \bibinfo{pages}{715} (\bibinfo{year}{2017}).

\bibitem[{\citenamefont{Ashcroft}(1968)}]{Ashcroft1968A}
\bibinfo{author}{\bibfnamefont{N.~W.} \bibnamefont{Ashcroft}},
  \bibinfo{journal}{Physical Review Letters} \textbf{\bibinfo{volume}{21}},
  \bibinfo{pages}{1748} (\bibinfo{year}{1968}).

\bibitem[{\citenamefont{Stadele and Martin}(2000)}]{Stadele2000A}
\bibinfo{author}{\bibfnamefont{M.}~\bibnamefont{Stadele}} \bibnamefont{and}
  \bibinfo{author}{\bibfnamefont{R.~M.} \bibnamefont{Martin}},
  \bibinfo{journal}{Physical Review Letters} \textbf{\bibinfo{volume}{84}},
  \bibinfo{pages}{6070} (\bibinfo{year}{2000}).

\bibitem[{\citenamefont{Cudazzo et~al.}(2008)\citenamefont{Cudazzo, Profeta,
  Sanna, Floris, Continenza, Massidda, and Gross}}]{Cudazzo2008A}
\bibinfo{author}{\bibfnamefont{P.}~\bibnamefont{Cudazzo}},
  \bibinfo{author}{\bibfnamefont{G.}~\bibnamefont{Profeta}},
  \bibinfo{author}{\bibfnamefont{A.}~\bibnamefont{Sanna}},
  \bibinfo{author}{\bibfnamefont{A.}~\bibnamefont{Floris}},
  \bibinfo{author}{\bibfnamefont{A.}~\bibnamefont{Continenza}},
  \bibinfo{author}{\bibfnamefont{S.}~\bibnamefont{Massidda}}, \bibnamefont{and}
  \bibinfo{author}{\bibfnamefont{E.~K.~U.} \bibnamefont{Gross}},
  \bibinfo{journal}{Physical Review Letters} \textbf{\bibinfo{volume}{100}},
  \bibinfo{pages}{257001} (\bibinfo{year}{2008}).

\bibitem[{\citenamefont{Szcz{\c{e}}{\'s}niak and
  Drzazga}(2013)}]{Szczesniak2013C}
\bibinfo{author}{\bibfnamefont{R.}~\bibnamefont{Szcz{\c{e}}{\'s}niak}}
  \bibnamefont{and} \bibinfo{author}{\bibfnamefont{E.~A.}
  \bibnamefont{Drzazga}}, \bibinfo{journal}{Solid State Sciences}
  \textbf{\bibinfo{volume}{19}}, \bibinfo{pages}{167} (\bibinfo{year}{2013}).

\bibitem[{\citenamefont{Yan et~al.}(2011)\citenamefont{Yan, Gong, and
  Liu}}]{Yan2011A}
\bibinfo{author}{\bibfnamefont{Y.}~\bibnamefont{Yan}},
  \bibinfo{author}{\bibfnamefont{J.}~\bibnamefont{Gong}}, \bibnamefont{and}
  \bibinfo{author}{\bibfnamefont{Y.}~\bibnamefont{Liu}},
  \bibinfo{journal}{Physics Letters A} \textbf{\bibinfo{volume}{375}},
  \bibinfo{pages}{1264} (\bibinfo{year}{2011}).

\bibitem[{\citenamefont{Szcz{\c{e}}{\'s}niak
  et~al.}(2012)\citenamefont{Szcz{\c{e}}{\'s}niak, Szcz{\c{e}}{\'s}niak, and
  Drzazga}}]{Szczesniak2012N}
\bibinfo{author}{\bibfnamefont{R.}~\bibnamefont{Szcz{\c{e}}{\'s}niak}},
  \bibinfo{author}{\bibfnamefont{D.}~\bibnamefont{Szcz{\c{e}}{\'s}niak}},
  \bibnamefont{and} \bibinfo{author}{\bibfnamefont{E.~A.}
  \bibnamefont{Drzazga}}, \bibinfo{journal}{Solid State Communications}
  \textbf{\bibinfo{volume}{152}}, \bibinfo{pages}{2023} (\bibinfo{year}{2012}).

\bibitem[{\citenamefont{Maksimov and Savrasov}(2001)}]{Maksimov2001A}
\bibinfo{author}{\bibfnamefont{E.~G.} \bibnamefont{Maksimov}} \bibnamefont{and}
  \bibinfo{author}{\bibfnamefont{D.~Y.} \bibnamefont{Savrasov}},
  \bibinfo{journal}{Solid State Communications} \textbf{\bibinfo{volume}{119}},
  \bibinfo{pages}{569} (\bibinfo{year}{2001}).

\bibitem[{\citenamefont{Szcz{\c{e}}{\'s}niak and
  Jarosik}(2009)}]{Szczesniak2009A}
\bibinfo{author}{\bibfnamefont{R.}~\bibnamefont{Szcz{\c{e}}{\'s}niak}}
  \bibnamefont{and} \bibinfo{author}{\bibfnamefont{M.~W.}
  \bibnamefont{Jarosik}}, \bibinfo{journal}{Solid State Communications}
  \textbf{\bibinfo{volume}{149}}, \bibinfo{pages}{2053} (\bibinfo{year}{2009}).

\bibitem[{\citenamefont{Szcz{\c{e}}{\'s}niak
  et~al.}(2014{\natexlab{a}})\citenamefont{Szcz{\c{e}}{\'s}niak, Duda, and
  Drzazga}}]{Szczesniak2014F}
\bibinfo{author}{\bibfnamefont{R.}~\bibnamefont{Szcz{\c{e}}{\'s}niak}},
  \bibinfo{author}{\bibfnamefont{A.~M.} \bibnamefont{Duda}}, \bibnamefont{and}
  \bibinfo{author}{\bibfnamefont{E.~A.} \bibnamefont{Drzazga}},
  \bibinfo{journal}{Physica C} \textbf{\bibinfo{volume}{501}},
  \bibinfo{pages}{7} (\bibinfo{year}{2014}{\natexlab{a}}).

\bibitem[{\citenamefont{McMahon and
  Ceperley}(2011{\natexlab{a}})}]{McMahon2011A}
\bibinfo{author}{\bibfnamefont{J.~M.} \bibnamefont{McMahon}} \bibnamefont{and}
  \bibinfo{author}{\bibfnamefont{D.~M.} \bibnamefont{Ceperley}},
  \bibinfo{journal}{Physical Review B} \textbf{\bibinfo{volume}{84}},
  \bibinfo{pages}{144515} (\bibinfo{year}{2011}{\natexlab{a}}).

\bibitem[{\citenamefont{McMahon and
  Ceperley}(2011{\natexlab{b}})}]{McMahon2011B}
\bibinfo{author}{\bibfnamefont{J.~M.} \bibnamefont{McMahon}} \bibnamefont{and}
  \bibinfo{author}{\bibfnamefont{D.~M.} \bibnamefont{Ceperley}},
  \bibinfo{journal}{Physical Review Letters} \textbf{\bibinfo{volume}{106}},
  \bibinfo{pages}{165302} (\bibinfo{year}{2011}{\natexlab{b}}).

\bibitem[{\citenamefont{Ashcroft}(2004)}]{Ashcroft2004A}
\bibinfo{author}{\bibfnamefont{N.~W.} \bibnamefont{Ashcroft}},
  \bibinfo{journal}{Physical Review Letters} \textbf{\bibinfo{volume}{92}},
  \bibinfo{pages}{187002} (\bibinfo{year}{2004}).

\bibitem[{\citenamefont{Drozdov et~al.}(2014)\citenamefont{Drozdov, Eremets,
  and Troyan}}]{Drozdov2014A}
\bibinfo{author}{\bibfnamefont{A.~P.} \bibnamefont{Drozdov}},
  \bibinfo{author}{\bibfnamefont{M.~I.} \bibnamefont{Eremets}},
  \bibnamefont{and} \bibinfo{author}{\bibfnamefont{I.~A.}
  \bibnamefont{Troyan}}, \bibinfo{journal}{arXiv: 1412.0460}
  (\bibinfo{year}{2014}).

\bibitem[{\citenamefont{Drozdov et~al.}(2015)\citenamefont{Drozdov, Eremets,
  Troyan, Ksenofontov, and Shylin}}]{Drozdov2015A}
\bibinfo{author}{\bibfnamefont{A.~P.} \bibnamefont{Drozdov}},
  \bibinfo{author}{\bibfnamefont{M.~I.} \bibnamefont{Eremets}},
  \bibinfo{author}{\bibfnamefont{I.~A.} \bibnamefont{Troyan}},
  \bibinfo{author}{\bibfnamefont{V.}~\bibnamefont{Ksenofontov}},
  \bibnamefont{and} \bibinfo{author}{\bibfnamefont{S.~I.}
  \bibnamefont{Shylin}}, \bibinfo{journal}{Nature}
  \textbf{\bibinfo{volume}{525}}, \bibinfo{pages}{73} (\bibinfo{year}{2015}).

\bibitem[{\citenamefont{Ma et~al.}(2015)\citenamefont{Ma, Duan, Li, Liu, Tian,
  Yu, Xu, Shao, Liu, and Cui}}]{Ma2015A}
\bibinfo{author}{\bibfnamefont{Y.}~\bibnamefont{Ma}},
  \bibinfo{author}{\bibfnamefont{D.}~\bibnamefont{Duan}},
  \bibinfo{author}{\bibfnamefont{D.}~\bibnamefont{Li}},
  \bibinfo{author}{\bibfnamefont{Y.}~\bibnamefont{Liu}},
  \bibinfo{author}{\bibfnamefont{F.}~\bibnamefont{Tian}},
  \bibinfo{author}{\bibfnamefont{H.}~\bibnamefont{Yu}},
  \bibinfo{author}{\bibfnamefont{C.}~\bibnamefont{Xu}},
  \bibinfo{author}{\bibfnamefont{Z.}~\bibnamefont{Shao}},
  \bibinfo{author}{\bibfnamefont{B.}~\bibnamefont{Liu}}, \bibnamefont{and}
  \bibinfo{author}{\bibfnamefont{T.}~\bibnamefont{Cui}},
  \bibinfo{journal}{arXiv:1511.05291}  (\bibinfo{year}{2015}).

\bibitem[{\citenamefont{Eliashberg}(1960)}]{Eliashberg1960A}
\bibinfo{author}{\bibfnamefont{G.~M.} \bibnamefont{Eliashberg}},
  \bibinfo{journal}{Soviet Physics JETP} \textbf{\bibinfo{volume}{11}},
  \bibinfo{pages}{696} (\bibinfo{year}{1960}).

\bibitem[{\citenamefont{Durajski et~al.}(2014)\citenamefont{Durajski,
  Szcz{\c{e}}{\'s}niak, and Duda}}]{Ania2014D}
\bibinfo{author}{\bibfnamefont{A.~P.} \bibnamefont{Durajski}},
  \bibinfo{author}{\bibfnamefont{R.}~\bibnamefont{Szcz{\c{e}}{\'s}niak}},
  \bibnamefont{and} \bibinfo{author}{\bibfnamefont{A.~M.} \bibnamefont{Duda}},
  \bibinfo{journal}{Solid State Communiations} \textbf{\bibinfo{volume}{195}},
  \bibinfo{pages}{55} (\bibinfo{year}{2014}).

\bibitem[{\citenamefont{Jarosik et~al.}(2015)\citenamefont{Jarosik, Wrona, and
  Duda}}]{Ania2015A}
\bibinfo{author}{\bibfnamefont{M.~W.} \bibnamefont{Jarosik}},
  \bibinfo{author}{\bibfnamefont{I.~A.} \bibnamefont{Wrona}}, \bibnamefont{and}
  \bibinfo{author}{\bibfnamefont{A.~M.} \bibnamefont{Duda}},
  \bibinfo{journal}{Solid State Communiations} \textbf{\bibinfo{volume}{219}},
  \bibinfo{pages}{1} (\bibinfo{year}{2015}).

\bibitem[{\citenamefont{Szcz{\c{e}}{\'s}niak
  et~al.}(2015)\citenamefont{Szcz{\c{e}}{\'s}niak, Jarosik, and
  Duda}}]{Ania2015B}
\bibinfo{author}{\bibfnamefont{R.}~\bibnamefont{Szcz{\c{e}}{\'s}niak}},
  \bibinfo{author}{\bibfnamefont{M.~W.} \bibnamefont{Jarosik}},
  \bibnamefont{and} \bibinfo{author}{\bibfnamefont{A.~M.} \bibnamefont{Duda}},
  \bibinfo{journal}{Advances in Condensed Matter Physics}
  \textbf{\bibinfo{volume}{2015}}, \bibinfo{pages}{969564}
  (\bibinfo{year}{2015}).

\bibitem[{\citenamefont{Szcz{\c{e}}{\'s}niak and Zem{\l}a}(2015)}]{Domin2015A}
\bibinfo{author}{\bibfnamefont{D.}~\bibnamefont{Szcz{\c{e}}{\'s}niak}}
  \bibnamefont{and} \bibinfo{author}{\bibfnamefont{T.~P.}
  \bibnamefont{Zem{\l}a}}, \bibinfo{journal}{Superconductor Science and
  Technology} \textbf{\bibinfo{volume}{28}}, \bibinfo{pages}{085018}
  (\bibinfo{year}{2015}).

\bibitem[{\citenamefont{Duda et~al.}(2017)\citenamefont{Duda, Szewczyk,
  Jarosik, Szcz{\c{e}}{\'s}niak, Sowi{\' n}ska, and
  Szcz{\c{e}}{\'s}niak}}]{Domin2017A}
\bibinfo{author}{\bibfnamefont{A.~M.} \bibnamefont{Duda}},
  \bibinfo{author}{\bibfnamefont{K.~A.} \bibnamefont{Szewczyk}},
  \bibinfo{author}{\bibfnamefont{M.~W.} \bibnamefont{Jarosik}},
  \bibinfo{author}{\bibfnamefont{K.~M.} \bibnamefont{Szcz{\c{e}}{\'s}niak}},
  \bibinfo{author}{\bibfnamefont{M.~A.} \bibnamefont{Sowi{\' n}ska}},
  \bibnamefont{and}
  \bibinfo{author}{\bibfnamefont{D.}~\bibnamefont{Szcz{\c{e}}{\'s}niak}},
  \bibinfo{journal}{Physica B: Condensed Matter,
  DOI:10.1016/j.physb.2017.10.107}  (\bibinfo{year}{2017}), \bibinfo{note}{{\it
  Characterization of the superconducting state in hafnium hydride under high
  pressure}}.

\bibitem[{\citenamefont{Szcz{\c{e}}{\'s}niak
  et~al.}(2014{\natexlab{b}})\citenamefont{Szcz{\c{e}}{\'s}niak, Durajski, and
  Szcz{\c{e}}{\'s}niak}}]{Domin2014A}
\bibinfo{author}{\bibfnamefont{D.}~\bibnamefont{Szcz{\c{e}}{\'s}niak}},
  \bibinfo{author}{\bibfnamefont{A.~P.} \bibnamefont{Durajski}},
  \bibnamefont{and}
  \bibinfo{author}{\bibfnamefont{R.}~\bibnamefont{Szcz{\c{e}}{\'s}niak}},
  \bibinfo{journal}{Journal of Physics: Condensed Mattter}
  \textbf{\bibinfo{volume}{26}}, \bibinfo{pages}{255701}
  (\bibinfo{year}{2014}{\natexlab{b}}).

\bibitem[{\citenamefont{Szcz{\c{e}}{\'s}niak
  et~al.}(2016)\citenamefont{Szcz{\c{e}}{\'s}niak, Durajski, and
  Szcz{\c{e}}{\'s}niak}}]{Domin2016A}
\bibinfo{author}{\bibfnamefont{R.}~\bibnamefont{Szcz{\c{e}}{\'s}niak}},
  \bibinfo{author}{\bibfnamefont{A.~P.} \bibnamefont{Durajski}},
  \bibnamefont{and}
  \bibinfo{author}{\bibfnamefont{D.}~\bibnamefont{Szcz{\c{e}}{\'s}niak}},
  \bibinfo{journal}{Physica Status Solidi B} \textbf{\bibinfo{volume}{253}},
  \bibinfo{pages}{538} (\bibinfo{year}{2016}).

\bibitem[{\citenamefont{Beach et~al.}(2000)\citenamefont{Beach, Gooding, and
  Marsiglio}}]{Beach2000A}
\bibinfo{author}{\bibfnamefont{K.~S.~D.} \bibnamefont{Beach}},
  \bibinfo{author}{\bibfnamefont{R.~J.} \bibnamefont{Gooding}},
  \bibnamefont{and}
  \bibinfo{author}{\bibfnamefont{F.}~\bibnamefont{Marsiglio}},
  \bibinfo{journal}{Physical Review B} \textbf{\bibinfo{volume}{61}},
  \bibinfo{pages}{5147} (\bibinfo{year}{2000}).

\bibitem[{\citenamefont{Bardeen
  et~al.}(1957{\natexlab{a}})\citenamefont{Bardeen, Cooper, and
  Schrieffer}}]{Bardeen1957A}
\bibinfo{author}{\bibfnamefont{J.}~\bibnamefont{Bardeen}},
  \bibinfo{author}{\bibfnamefont{L.~N.} \bibnamefont{Cooper}},
  \bibnamefont{and} \bibinfo{author}{\bibfnamefont{J.~R.}
  \bibnamefont{Schrieffer}}, \bibinfo{journal}{Physical Review}
  \textbf{\bibinfo{volume}{106}}, \bibinfo{pages}{162}
  (\bibinfo{year}{1957}{\natexlab{a}}).

\bibitem[{\citenamefont{Bardeen
  et~al.}(1957{\natexlab{b}})\citenamefont{Bardeen, Cooper, and
  Schrieffer}}]{Bardeen1957B}
\bibinfo{author}{\bibfnamefont{J.}~\bibnamefont{Bardeen}},
  \bibinfo{author}{\bibfnamefont{L.~N.} \bibnamefont{Cooper}},
  \bibnamefont{and} \bibinfo{author}{\bibfnamefont{J.~R.}
  \bibnamefont{Schrieffer}}, \bibinfo{journal}{Physical Review}
  \textbf{\bibinfo{volume}{108}}, \bibinfo{pages}{1175}
  (\bibinfo{year}{1957}{\natexlab{b}}).

\bibitem[{\citenamefont{Carbotte and Marsiglio}(2003)}]{Carbotte2003A}
\bibinfo{author}{\bibfnamefont{J.~P.} \bibnamefont{Carbotte}} \bibnamefont{and}
  \bibinfo{author}{\bibfnamefont{F.}~\bibnamefont{Marsiglio}}, in
  \emph{\bibinfo{booktitle}{The Physics of Superconductors edited by K. H.
  Bennemann and J. B. Ketterson}} (\bibinfo{publisher}{Springer Berlin
  Heidelberg}, \bibinfo{year}{2003}).

\end{thebibliography}
\end{document}